\documentclass[%
 reprint,
 amsmath,amssymb,
 $aps,
prb,
]{revtex4-1}

\usepackage{graphicx} % Include figure files
\usepackage{bm}
\usepackage{txfonts}
\usepackage{mathrsfs}
\usepackage{amsmath}
\usepackage{xcolor}

\usepackage{subcaption}

\def\b0{{\mathbf 0}}

\def\b0{{\mathbf 0}}

\def\beq{\begin{equation}}
\def\eeq{\end{equation}}

\begin{document}

\title{Criticality of the $O(2)$ model with cubic anisotropies from nonperturbative renormalization} 
\author{Andrzej Chlebicki}
\affiliation{Institute of Theoretical Physics, Faculty of Physics, University of Warsaw, Pasteura 5, 02-093 Warsaw, Poland}
%\affiliation{Max-Planck-Institute for Solid State Research, Heisenbergstr.\ 1, D-70569 Stuttgart, Germany}
%
\author{Pawel Jakubczyk}
\affiliation{Institute of Theoretical Physics, Faculty of Physics, University of Warsaw, Pasteura 5, 02-093 Warsaw, Poland} 
\date{\today}
\begin{abstract}
We study the $O(2)$ model with $\mathbb{Z}_4$-symmetric perturbations within the framework of nonperturbative renormalization group (RG) for spatial dimensionality $d=2$ and $d=3$. In a unified framework we resolve the relatively complex crossover behavior emergent due to the presence of multiple RG fixed points. In $d=3$ the system is controlled by the $XY$, Ising, and low-$T$ fixed points in presence of a dangerously irrelevant anisotropy coupling $\lambda$. In $d=2$ the anisotropy coupling is marginal and the physical picture is governed by  the interplay between two distinct lines of RG fixed points, giving rise to nonuniversal critical behavior; and an isolated Ising fixed point. In addition to inducing crossover behavior in universal properties, the presence of the Ising fixed point yields a generic, abrupt change of critical temperature at a specific value of the anisotropy field.

  %In $d=3$ the anisotropy coupling is 

\end{abstract}


\maketitle

%%%%%%%%%%%%%%%%%%%%%%%%%%%%%%%%%%%%%%%%%%%%%%%%%%%%%%%%%%%%%%%%%%%%%%%%%%

\section{Introduction}
Universality is a hallmark of second order phase transitions\cite{Goldenfeld_book, Zinn_Justin_small} and its explanation may be viewed as a major success of renormalization group  theory. There are however well known cases, where the critical properties are actually not universal. Commonly recognized examples include systems with long-ranged interactions,\cite{long_range_Les_Houches} interfacial unbinding transitions in $d=3$,\cite{Parry_2009} some spin glass transitions,\cite{Bernardi_1996} or the eight vertex model.\cite{Baxter_1971} In the RG framework universality emerges as consequence of the existence of isolated fixed points associated with a unique set of relevant perturbations characterized by their scaling dimensions. One way of avoiding the emergence of universality appears when  the fixed point features a marginal operator. In such situations the flow diagram may exhibit a line of fixed points and the critical exponents may vary continuously depending on the system parameters and the thermodynamic fields. 

In this paper we address the classical $O(2)$ model with  perturbations which explicitly break the symmetry from $O(2)$ to $\mathbb{Z}_4$. From the perspective of the general theory of continuous phase transitions this system is interesting and somewhat unusual both in dimensionality $d=3$ and $d=2$. In the former case the anisotropy field $\lambda$ acts as an irrelevant operator and the critical behavior is governed by the standard $XY$ (Wilson-Fisher) fixed point. However, $\lambda$ gaps the Goldstone mode in the low-$T$ phase and gives rise to the emergence of one additional length scale which diverges at the critical point. This leads to the effect of distinct susceptibility and correlation length exponents depending on the side from which the phase transition is approached. In dimensionality $d=2$ the anisotropy stabilizes the long-range ordered phase and  is known to be a marginal perturbation at the Kosterlitz-Thouless (KT) phase transition occurring at $\lambda=0$. This leads to the appearance of two additional fixed-point lines emerging towards positive and negative values of $\lambda$ from the vicinity of the endpoint of the KT fixed point line located at $\lambda=0$ and  $T<T_{KT}$. The existence of these lines renders the exponents characterizing the transition at $\lambda\neq 0$ nonuniversal. The 2-dimensional $O(2)$ model with $\mathbb{Z}_4$ anisotropies is also of high experimental relevance in a number of contexts. For a broad and useful exposition we refer to Ref.~\onlinecite{Taroni_2008}.
In addition, both for $d=2$ and $d=3$ in the critical regime the system is equivalent to the Ising model for specific choices of the model couplings. These features lead to reach crossover phenomena and in $d=2$ raise the interesting question concerning the relation between the Ising fixed point and the above-mentioned fixed-point line. The purpose of the present paper is to analyze this interplay and to understand the evolution of the RG flow diagram of the system upon lowering the dimensionality from $d=3$ to $d=2$ in a nonperturbative RG framework, which allows for an (approximate) capturing of the essential features both in $d=2$ and $d=3$ and resolving the rich crossover phenomena. 

The paper is structured as follows: in Sec.~II  we set the context for the present study by giving an overview of the most important results obtained earlier for this system within different approaches both in $d=3$ and $d=2$ . In Sec.~III we present the formulation of the model applied in our analysis, while in Sec.~IV we  review the nonperturbative RG framework which is followed by derivation and discussion of the RG flow equations. Certain features of the system can be elucidated from the analysis of the asymptotic forms of flow equations. In Sec.~V we present our results obtained from the numerical integration of the RG equations in a simple truncation varying dimensionality between $d=2$ and $d=3$. 
%In Sec.~VI we readdress the system going beyond the simple truncation.  
We give a summary of our results in Sec.~VI.   

\section{Overview of key earlier results}
In their seminal paper\cite{Jose_1977} Jos\'e \textsl{ et al } analyzed the two-dimensional $XY$ model to which they added $p$-fold anisotropy terms of the form $\sum_{\bf{i}} h_p \cos (p\theta_{\bf i})$, where $\bf i$ enumerates the lattice sites, $\theta_{\bf i}$ is the corresponding angular variable, while $p$ characterizes the anisotropy field. Relying on the Villain approximation they investigated the relevance of the symmetry-breaking perturbations $h_p$ along the line $h_p=0$ in the low-$T$ (KT) phase. They found 
\beq
4\leq 2\pi K_{eff}\leq\frac{1}{4}p^2 
\label{T_KT}
\eeq
as the condition for the stability of the spin-wave theory with respect to a $p$-fold degenerate perturbation and vortices. In the above $K_{eff}$ is the renormalized stiffness.  The first inequality yields the usual KT instability temperature. The relation (\ref{T_KT}) implies the existence of a KT-like phase for a regime of temperatures also for nonvanishing anisotropy field $h_p$ provided $p>4$. For $p=4$ this regime becomes degenerate (for given value of $h_4$) to a single point. The emergent picture yields (at least for small $h_4$) a marginal operator associated with the presence of the anisotropy. As a result Jos\'e \textsl{ et al } predicted nonuniversal exponents along the transition line in the $(T-h_4)$ plane. Significantly later the two-dimensional $XY$ model with cubic anisotropies was addressed by extensive Monte-Carlo (MC) simulations\cite{Rastelli_2004_1, Rastelli_2004_2} performed on the square lattice $XY$ model. These studies confirmed the nonuniversal character of the critical exponents along the transition line for $h_4\neq 0$, but, interestingly, disagreed with  Ref.~\onlinecite{Jose_1977} on the very structure of the phase diagram, pointing towards the stability of the KT phase for sufficiently low $h_4$ and yielding a picture similar to that obtained for higher values of $p$. Another controversial aspect concerns the range of conceivable values of critical exponents for this class of models. In particular, Ref.~\onlinecite{Jose_1977} predicted the divergence of the order parameter exponent $\beta$ for vanishing $h_4$ ($\beta\sim h_4^{-1}$), so that in principle the set of allowed values of $\beta$ is unbounded from above. However,\cite{Taroni_2008} experimentally observed values fall within a window $\beta\in [1/8,0.23]$ and point towards the existence of an upper bound on $\beta$. Ref.~ \onlinecite{Taroni_2008} attributed this to  suppression of the predicted nonuniversal criticality by finite-size scaling properties of the 2-dimensional XY model and identified the value $\beta\approx 0.23$ as the effective critical exponent of the pure XY model due to finite system size. The length scale of the onset of the true critical singularities at small $h_4$ is then extremely large and at realistic system sizes, they become overshadowed by the KT-type scaling.
 Our present study is consistent with this point of view.

For $d=3$ the XY model with cubic anisotropies was (presumably) first addressed by Aharony\cite{Aharony_1973} within the $\epsilon$ expansion. The perturbative RG approach for this system was later significantly developed by Carmona \textsl{ et al}.\cite{Carmona_2000} In particular, they identified the fixed points present in the phase diagram and investigated their stability upon varying the number of field components. 
 In a relatively recent study\cite{Leonard_2015} L\'eonard and Delamotte employed nonperturbative RG to address a generic mechanism leading to distinct exponents characterizing the critical behavior approaching the transition from high and low temperatures. They analyzed the $O(2)$ model with discrete anisotropies as an illustrative example of this phenomenon, which   had long before been observed by Nelson.\cite{Nelson_1976} In the present work we employ a framework similar to Ref.~\onlinecite{Leonard_2015} to investigate the evolution of the system as the dimensionality of the system is reduced from $d=3$ anticipating the appearance of  nonuniversal critical features for dimensionality $d$ approaching 2.\cite{Codello_2013}

\section{Model}
We consider the $d$-dimensional classical $O(2)$ model supplemented with a cubic anisotropy term 
 \begin{align}
&S[\phi] = \int d \bm{r} \left[\frac{u_0}{8}(2\rho - \alpha_0^2)^2 + \frac{\lambda_0}{2} \tau + \frac{1}{2} |\nabla\phi|^2  \right]. 
\label{Bare_action}
    \end{align}    
with a (real) two-component order-parameter field $\phi=\phi(\bm{r})=(\phi_1(\bm{r}), \phi_2(\bm{r}))$. The $O(2)$ invariant $\rho$ and the $\mathbb{Z}_4$ invariant $\tau$ are defined as follows 
 \beq
 2\rho = |\phi|^2 =\phi_1^2 +\phi_2^2\;,\;\;\;\;\; \tau = \phi_1^2\phi_2^2\;. 
 \label{Invariants}
 \eeq 
 The uniform contribution involves two quartic couplings: the usual $O(2)$ coupling $u_0$ and the anisotropy strength $\lambda_0$. We observe that Eq.~(\ref{Bare_action}) is invariant under a rotation of the field $\phi$ by $\frac{\pi}{4}$ followed by the transformation $u_0' = u_0+\lambda_0$, $\lambda_0'=-\lambda_0$, $\alpha_0'^2 = \alpha_0^2 \frac{u_0'}{u_0}$. In consequence (for $|\lambda_0|<u_0$), we can assume $\lambda_0>0$ without loss of generality. 
  We concentrate on the situation with symmetry-breaking at mean-field (MF) level, where $\alpha_0^2>0$. The units are chosen so that the coefficient of the gradient term in Eq.~(\ref{Bare_action}) is $\frac{1}{2}$. Note that our definition of $\tau$ differs from that of Refs.~\onlinecite{Carmona_2000, Leonard_2015} which adopted $\tau=\phi_1^4+\phi_2^4$. The two definitions correspond to picking reference frames related by a rotation. 
  One advantage of the present choice is that the minima of the effective potential $U(\rho,\tau)=\frac{u_0}{8}(2\rho - \alpha_0^2)^2 + \frac{\lambda_0}{2} \tau$ are located on the lines $\phi_1=0$ and $\phi_2=0$ as illustrated in Fig.~1. 
In the absence of the anisotropy ($\lambda_0=0$) Eq.~(\ref{Bare_action}) reduces to the standard $O(2)$ model, which may exhibit long-range order only for $d>2$.\cite{Mermin_1966} The algebraic KT phase occurs  for $d=2$ if $\alpha_0^2$ is sufficiently large.
 As soon as $\lambda_0>0$, the $O(2)$ degeneracy of the ground-state becomes reduced to discrete fourfold degeneracy as illustrated in Fig.~1, and the occurrence of true long-range order is then permissible down to dimensionality $d=1^+$.
 \begin{figure}
  \begin{subfigure}[b]{0.23\textwidth}
    \includegraphics[width=\textwidth]{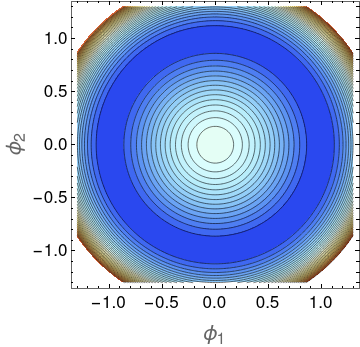}
  %  \caption{Picture 1}
   % \label{fig:1}
  \end{subfigure}
  \begin{subfigure}[b]{0.23\textwidth}
    \includegraphics[width=\textwidth]{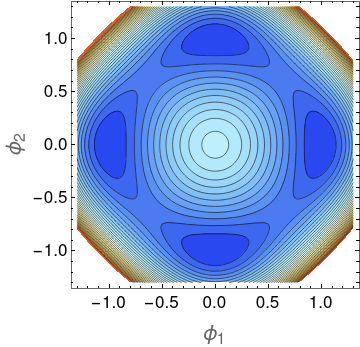}
   % \caption{Picture 2}
    %\label{fig:2}
  \end{subfigure} 
  \caption{(Color online) Contour plots of the effective potential for the pure $O(2)$ model (left panel) and the $O(2)$ model with a $\mathbb{Z}_4$ anisotropy $\tau$ given  by Eq.~(\ref{Invariants}) (right panel). In the latter case the ground state is fourfold degenerate with the minima located at the $\phi_1=0$ and $\phi_2=0$ axes.   }
\end{figure}
%\begin{figure}[ht]
%\begin{center}
%\label{Fig1}
%\includegraphics[width=6cm]{Fig_1.pdf}
%\caption{(Color online) Contour plot of the effective potential of the $O(2)$ model with a $\mathbb{Z}_4$ anisotropy $\tau$ given  by Eq.~(\ref{Invariants}). The ground state is fourfold degenerate with the minima located at the $\phi_1=0$ and $\phi_2=0$ axes.  }
%\end{center}
%\end{figure} 
 \section{Nonperturbative RG approach}
We employ the 1-particle irreducible variant of non-perturbative renormalization theory to investigate the model defined by Eq.~(\ref{Bare_action}). This is very suitable for the present problem, allowing for capturing the distinct scaling regimes and associated crossover behavior as the system is evolved from the short to long observation scales. We consider the flowing effective action $\Gamma_k[\phi]$, which interpolates between the bare action [Eq.~(\ref{Bare_action})] and the thermodynamic free energy $G[\phi]$ as the flow parameter $k$, implemented as an infrared momentum cutoff, is reduced from the UV cutoff scale $k=\Lambda$ towards zero. The quantity $\Gamma_k[\phi]$ also acts as the generating functional for 1-particle irreducible vertex functions in presence of the IR regulator. The idea to continuously integrate fluctuations out of the partition function via the RG flow is implemented by deforming the propagators so that the fluctuation modes with momentum lower then the cutoff scale ($q<k$) acquire an artificial mass of order $k^2$. The scale $k$ may then be continuously varied leading to the evolution of $\Gamma_k[\phi]$ governed by the Wetterich equation\cite{Wetterich_1993}
\beq
\partial_t\Gamma_k[\phi] = \frac{1}{2}\textrm{Tr}\left\{\partial_t R_k \left[\Gamma_k^{(2)}[\phi] +R_k\right]^{-1}\right\}\;, 
\label{Wetterich}
\eeq
where $\Gamma_k^{(2)}[\phi]$ is the second field derivative of $\Gamma_k[\phi]$ and $R_k$ denotes the cutoff function added to the inverse propagator to damp the modes with $q<k$. The trace sums over momenta and  field components, while $t=\log(k/\Lambda)$. The Wetterich framework was successfully used in a diversity of contexts over the last years (for reviews see e.g.~\onlinecite{Berges_2002, Pawlowski_2007, Kopietz_book, RG_book, Metzner_2012}).

The functional differential equation given by Eq.~(\ref{Wetterich}) can hardly ever be solved exactly. Here we resort to the approximation scheme known as the derivative expansion.\cite{Berges_2002, Canet_2003, Delamotte_2004, RG_book}  This  classifies the symmetry-allowed terms in $\Gamma_k[\phi]$ according to the number of derivatives (or momentum powers in Fourier space). We parametrize $\Gamma_k[\phi] $ as 
\beq
 \Gamma_k[\phi] = \int d\bm{r}\left\{U_k(\rho,\tau) + \frac{1}{2}Z_k|\nabla\phi|^2+\frac{1}{8}Y_k(\nabla |\phi|^2)^2  \right\}\;, 
 \label{Gamma}
 \eeq
 where we neglect the dependence of $Z_k$ and $Y_k$ on $\rho$ and $\tau$.
  In the low-temperature phase the the effective potential features four degenerate minima corresponding to the ground states. In the present simple truncation we parametrize $U_k(\rho,\tau)$ by expanding around one of them up to the lowest sensible order 
 \beq
 U_k(\rho,\tau) = \frac{u_k}{2}(\rho-\rho_{0,k})^2+\frac{\lambda_k\tau}{2}\;,  
 \label{U_truncation}
 \eeq
 where $\rho_{0,k}=\frac{1}{2}\alpha_k^2$ describes the flowing distance of the minimum from the origin.
This approximation level allows for a degree of analytical understanding of the flow and for a rather straightforward resolution of the crossover behavior by direct integration of the RG equations. %We shall discuss results from a higher-order truncation later in Sec.~VI.

Within the approximation defined by Eqs.~(\ref{Gamma}) and (\ref{U_truncation}) the problem of integrating Eq.~(\ref{Wetterich}) becomes reduced to the analysis of a set of five coupled integro-differential equations governing the flow of the couplings $\{\alpha_k^2, u_k, \lambda_k, Z_k, Y_k\}$. In $d=3$ an analogous approach was employed in Ref.~\onlinecite{Leonard_2015} with major focus on the hexagonal rather than cubic anisotropy. In that work the effective potential expansion [Eq.~\ref{U_truncation}] was pushed to higher order. This (at least for the pure $O(N)$ models) allows for obtaining results convergent to the complete derivative expansion for $d$ close to 3 (but not for $d$ approaching 2). On the other hand, we retain the $Y_k$ coupling in the present treatment. This is known\cite{Jakubczyk_2017_2} to significantly improve the quality of the approximation for $d=2$. On the technical level, accounting for $Y_k$ distinguishes between momentum-dependent terms in the longitudinal and transverse directions. The present functional RG truncation does not explicitly invoke vortices.\cite{Kosterlitz_1973} In consequence, for $\lambda=0$ the line of fixed points in the low-$T$ phase is recovered only approximately in the form of quasi-fixed points unless the regulator is fine tuned.\cite{Jakubczyk_2014} This issue was first investigated in Refs.~\onlinecite{Graeter_1995, Gersdorff_2001}, and recently was revisited in a sequence of works\cite{Jakubczyk_2014, Defenu_2017, Krieg_2017, Jakubczyk_2017_2} providing insightful understanding of the flow and proposing avenues for improvements. In particular, Ref.\onlinecite{Krieg_2017} showed how the hierarchy of functional RG flow equations can be reduced to the Kosterlitz-Thouless RG equations. 
On the other hand, the functional RG framework serves as a particularly convenient tool in situations involving rich crossover phenomena.\cite{Strack_2009, Jakubczyk_2010, Leonard_2015, Lammers_2016, Debelhoir_2016, Rancon_2017} and (at the cost of working at truly functional level) is capable of computing also nonuniversal aspects of specific microscopic models.\cite{Machado_2010, Jakubczyk_2016}

With the parametrization specified by Eqs.~(\ref{Gamma}) and (\ref{U_truncation}) the regularized propagator is given by 
\begin{equation}
\left[\Gamma_k^{(2)}\right]^{-1}_{i,j}=G_{i}(\bm{q}) = \frac{\delta_{i,j}}{m_{i}^2 + Z_{i} \bm{q}^2 + R_k(\bm{q}^2)}\;, 
\end{equation}
where $i,j\in\{1,2\}$. In our convention $i=1$ corresponds to the longitudinal and $i=2$ to the transverse mode. The flowing masses and $Z$-factors are given by
\begin{alignat}{2}
&m_1^2 = u \alpha^2\;, \hspace{2cm} &&m_2^2 = \lambda \alpha^2\;, \\
&Z_1 = Z + Y\alpha^2\;, &&Z_2 = Z \nonumber\;.
\end{alignat}
We suppressed the $k$-dependencies in our notation for clarity. We observe that the essential role of the anisotropy coupling $\lambda$ is to give the transverse mode a mass. At the level of higher-order vertex functions one can easily show that $\lambda$ will influence the vertices with an even number of transverse legs. We also introduce the single-scale propagators
\beq	
G_{i}'(\bm{q}) = -G_{i}^2 \partial_t R\;.
\eeq 
and the operator $D_t=(\partial_t R) \partial_R$\;.
\subsection{Flow equations}
The derivation of the flow equations follows a standard procedure (see e.g. Ref.~\onlinecite{Berges_2002}). The flow of $\alpha^2$ is extracted from the condition $\frac{d}{dt}U'(\rho,\tau)|_{\rho=\alpha^2/2,\tau=0}  =0$. The flow of the masses follows from first differentiating Eq.~(\ref{Wetterich}) twice with respect to the field $\phi_i$, and evaluating at a uniform field configuration, thus deriving a flow equation for the longitudinal and transverse components of $\Gamma_k^{(2)}$. Subsequently, taking the limit $q\to 0$ yields the flow of the masses $m_i^2$. The flow of $Z_1$ and $Z_2$ is obtained from expanding the flow of $\Gamma_k^{(2)}$ in the external momentum and picking the coefficients of order $q^2$. The resulting flow equation for $\alpha^2$ reads: 
\beq
\partial_t{\alpha^2} = -\frac{1}{u} \int_{\bm{q}}\left[(u+2U(\bm{q}))G_1'(\bm{q}) +(u+2\lambda)G_2'(\bm{q})\right]\;,
\eeq
while the flow of the masses is given by
 \begin{align}
\partial_t{m^2_1} =& -  \int_{\bm{q}}\left[(u+2U(\bm{q})) G_1'(\bm{q}) + (u+2\lambda) G_2'(\bm{q}) \right] \\
&-  \alpha^2 \int_{\bm{q}}\left[(u + 2U(\bm{q}))^2 G'_1(\bm{q})G_1(\bm{q}) + (u+ 2\lambda)^2 G'_2(\bm{q})G_2(\bm{q})\right], \nonumber \\
\partial_t{m^2_2} =& -\frac{ m_2^2}{m_1^2} \int_{\bm{q}}\left[(u+2U(\bm{q}))G_1'(\bm{q}) +(u+2\lambda)G_2'(\bm{q})\right] \\
&-  3m_2^2 \int_{\bm{q}}(2U(\bm{q}) + \lambda) D_t (G_1(\bm{q})G_2(\bm{q}))\;. \nonumber
\end{align}
 We introduced $\int_{\bm{q}}=\int\frac{d\bm{q}}{(2\pi)^d}$ and $U(\bm{q})=u+Y\bm{q}^2$. We relegate the expressions for the flow of $Z$-factors to the appendix. Upon putting $\lambda=0$ we recover the flow equations well studied for the $O(2)$ model.\cite{Berges_2002, Dupuis_2011, RG_book, Jakubczyk_2017_2} The presence of the mass of the transverse mode underlies the mechanism discussed in Ref.~\onlinecite{Leonard_2015} responsible for generating the unequal critical exponents in the low- and high-temperature phases.  There is a clear asymmetry between the properties of the flow of the two masses. If $m_2^2=0$ at some scale, it will never be generated. The converse is obviously not true. It is worth noting that when putting $\lambda=u$, $Y=0$ the flow equations of the longitudinal and transverse masses become identical and the property $\lambda=u$ remains conserved by the flow. Moreover, the resulting equations are equivalent to the corresponding flow of the Ising [$O(1)$] universality class (at the same level of approximation). This means, that if (initially) $\lambda=u$, the phase transition is bound to fall in the Ising universality class. This conclusion holds true also beyond the present approximation level and at any dimensionality. This is also consistent with the results of Ref.~\onlinecite{Carmona_2000}, which identified an Ising-type fixed point in the flow diagram. Another simplification occurs in the high-anisotropy limit $\lambda\to\infty$, where the transverse fluctuations are suppressed. It is then easily shown that the flow of $\alpha$, $u$ and $Z_{\sigma}$ is equivalent that corresponding to the $O(1)$ model. We note however that the present truncation level is probably not quite reliable to address the regime of strong anisotropies and postpone this case to future studies. In $d=2$ and for $\lambda=0$ the IR scaling of the Goldstone propagator follows $G_2\sim k^{2-\eta}$. Its coupling to the longitudinal mode gives rise to similar asymptotic IR scaling of the longitudinal mode $G_1$ as well. This behavior becomes suppressed by the presence of the anisotropy $\lambda$, which is a relevant coupling for low $T_{eff}=\alpha_0^{-2}$ and destabilizes the KT-like phase towards formation of long-ranged order. For small $\lambda_0$ this effect may occur only at asymptotically low scales. For $T_{eff}$ approaching the critical value (and small $\lambda_0$) the flow exhibits very slow evolution controlled by KT (quasi)fixed point line before crossing over to another regime, controlled by another (quasi)fixed point line. This is demonstrated by the numerical solution presented in the next section.  
 In the following we will work with the dimensionless variables defined as 
\begin{align}
\label{rescaling}
\tilde \kappa = k^{2-d} Z \alpha^2, \hspace{0.1cm} \tilde u = k^{d-4}  Z^{-2} u, \hspace{0.1cm} \tilde \lambda = k^{d-4}  Z^{-2} \lambda, \hspace{0.1cm} \tilde Y = k^{d-2} Z^{-1} Y\;,
\end{align} 
in terms of which fixed-point behavior is transparent. 
\subsection{Note on the scaling laws}
Presence of the dangerously irrelevant coupling $\lambda$ influences the scaling laws relating the critical exponents which characterize the correlation functions. We rederived these relations following the line of reasoning of Ref.~\onlinecite{Leonard_2015} (see also \onlinecite{Okubo_2015}). We obtained 
\begin{align}
\label{full_scaling}
&\nu' = \nu (1 + y/2)\;, \\
&\gamma^+ = \nu(2-\eta)\;, \nonumber\\
&\gamma_T = \gamma^+ + \nu y\;, \nonumber\\
&\gamma_L = \gamma^+ + \nu y \frac{4-d-\eta}{2}\;, \nonumber
\end{align}
where $\gamma_T$, $\gamma_L$ are the critical exponents for the transverse and longitudinal susceptibilities, $\gamma^+$ controls the susceptibility divergence when the critical point is approached from the high-temperature phase,  $\eta$ is the anomalous dimension, while $y$ is the scaling exponent for $\lambda$ at the $XY$ fixed point (i.e. $\tilde{\lambda}\sim k^y$). The quantities $\nu$ and $\nu'$ denote the critical exponent for the longitudinal and transverse correlation lengths. Eq.~(\ref{full_scaling}) are valid for arbitrary anisotropies (not necessarily cubic) and for any $d$ provided $\lambda$ remains an irrelevant variable. 
 The last of the relations of Eq.~(\ref{full_scaling}) differs from the form given in Ref.~\onlinecite{Leonard_2015} by the $\eta$ present in our expression.    
 
 The modification of the scaling laws as compared to the pure $O(N)$ case may be traced back to the fact that there exist two large length scales associated to the two directions in the field space. They are determined by the RG scales at which the flow departs from the $XY$ and the low-temperature fixed points respectively.\cite{Leonard_2015} The latter scale is always infinite in the absence of anisotropies. Observe that,\cite{Leonard_2015}  quite counterintuitively, the difference between the susceptibility exponents grows upon increasing $y$ so that the less relevant the perturbation, the larger the difference between the critical indices and the deviation from the isotropic case.

\section{Integration of the RG flow}
We now present the results obtained from numerical integration of the flow equations in the simple truncation defined in Sec.~IVA. We implement the exponential cutoff\cite{Berges_2002} 
\begin{equation}
    R_k(\bm q^2) =  A \frac{Z_k \bm q^2}{e^{\bm q^2/k^2}-1}\;,
    \end{equation}
where we take $A=2$.\cite{Jakubczyk_2014} We solve the flow equations discussed in Sec.~IVA with the initial condition provided by Eq.~(\ref{Bare_action}). We identify the ordered phase by the condition $\lim_{k\to 0}\alpha^2>0$. On the other hand, for the high-temperature phase the flowing order parameter $\alpha$ reaches zero at a finite value of $k$.   For a given fixed initial value of $u=u_0$ we evaluate the critical line $\lambda_{0,u}(\alpha_0)$. The quantity $\alpha_0^{-2}=T_{eff}$ may be understood as an effective temperature of the model. From the analysis of the flow equations in Sec.~IVA it is clear that precisely one point (corresponding to $\lambda_0=u_0$) on the continuum of the critical line is a representative of the Ising universality class. Except for this the critical behavior in $d=3$ is controlled by the Wilson-Fisher fixed point with the dangerously irrelevant coupling $\lambda$, while in $d=2$ one expects the fixed-point line discussed in Ref.~\onlinecite{Jose_1977}. It is perhaps of particular interest to understand the relation between this fixed-point line and the distinct Ising fixed point.\subsection{Results in $d=3$}
In Fig.~1 we present the phase diagrams computed for the model defined by Eq.~(\ref{Bare_action}) in the simple truncation in $d=3$ for a sequence of values of $u_0$. 
\begin{figure}[ht]
\begin{center}
\label{Fig2}
\includegraphics[width=8.5cm]{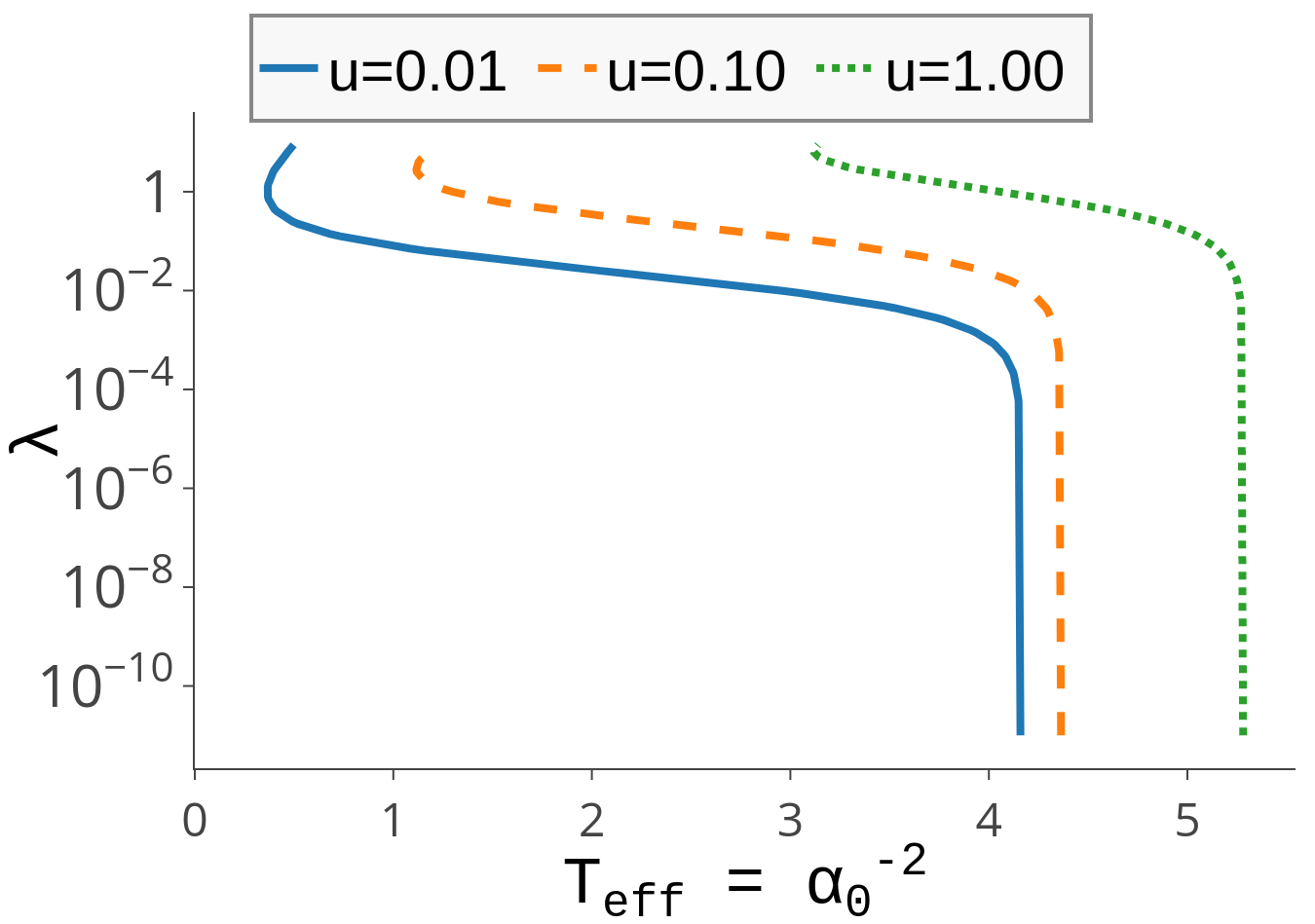}
\caption{(Color online) The critical line $\lambda_{0,u}(\alpha_0)$ computed for a sequence of values of $u_0$ in $d=3$. For low anisotropies the critical value of $T_{eff}$ is practically constant. The sudden drop of $T_{eff}$ occurs at $\lambda_0\approx u_0$, where the transition is in the Ising universality class. The same feature occurs for $d=2$ (see Fig.~6).  }
\end{center}
\end{figure} 
A striking (and counterintuitive) feature is the sudden drop of $T_{eff}$ at larger values of $\lambda_0$, precisely corresponding to $\lambda_0= u_0$, where the transition is in the Ising universality class. This feature of the phase diagram is understood from the flow by observing that the position of the Ising and $XY$ fixed points in the flow diagram are very different. The pronounced  decrease of $T_{eff}$ upon rising $\lambda_0$ towards $u_0$ is then a consequence of continuity of the flow. Fig.~2 provides an example manifestation of crossover of a \emph{ nonuniversal} thermodynamic quantity (the critical line) due to an interplay between two RG fixed points. The magnitude of this effect depends strongly on the value of $u_0$. Another manifestation of this crossover behavior is identified by inspecting the critical exponents. As an illustrative example in Fig.~3 we plot the transverse mass as function of $T^c_{eff}-T_{eff}$ and observe the crossover of the corresponding critical exponent $\gamma_T$ between two values related to the Ising and $XY$ fixed points.  
\begin{figure}[ht]
\begin{center}
\label{}
\includegraphics[width=8.5cm]{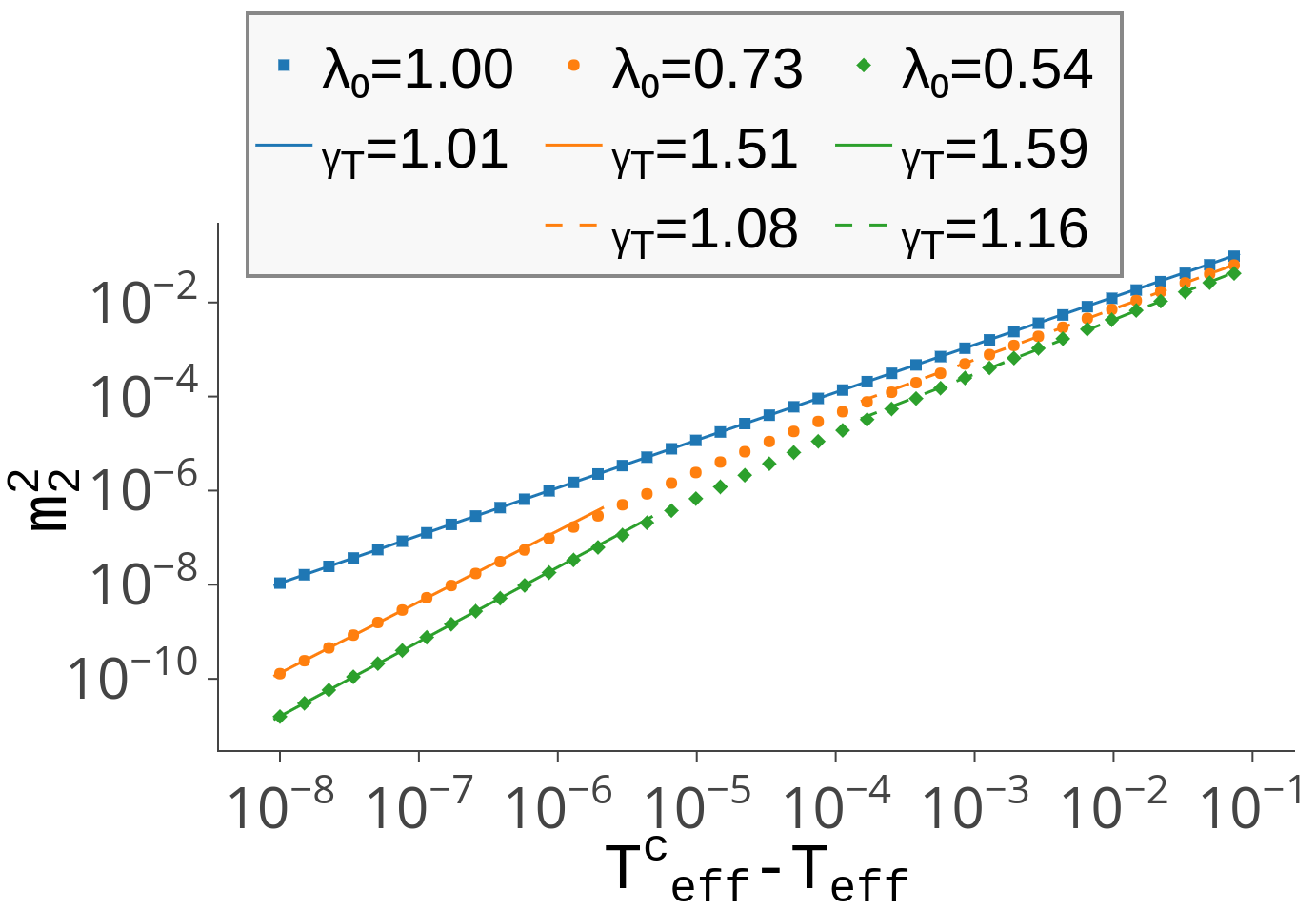}
\caption{(Color online) Renormalized transverse mass $m_2$ (inverse transverse susceptibility) as function of $T^c_{eff}-T_{eff}$ for a sequence of values of $\lambda_0$ and $u_0=1$. Upon varying  $\lambda_0$ away from $\lambda_0=u_0$ the corresponding exponent $\gamma_T$ crosses over from the Ising behavior (the upper, blue curve) to the XY value.}
\end{center}
\end{figure} 

Crossover behavior manifests itself also by varying the observation scale. For the present situation we exemplify this in Fig.~4 by plotting the anomalous dimension $\eta$ versus the cutoff scale. It exhibits two scaling plateaus describing the Ising and XY universality classes. The specific values of $\eta$ are off the accurate ones by a factor of order 2, which is due to truncation (see e.g. Ref.~\onlinecite{Strack_2009} for comparison). 
\begin{figure}[ht]
\begin{center}
\label{}
\includegraphics[width=8.5cm]{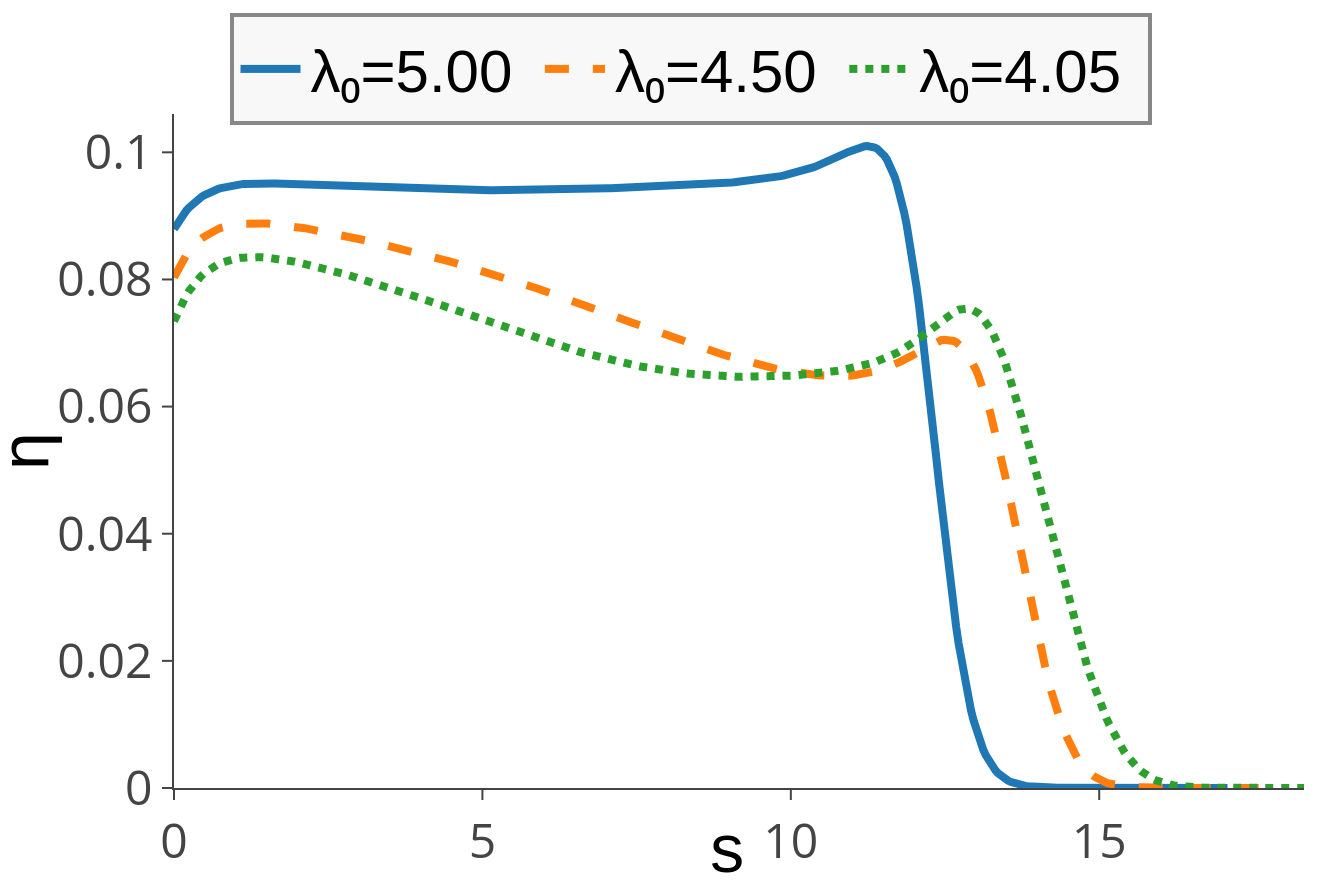}
\caption{(Color online)  The flowing anomalous dimension $\eta$ as function of the renormalization scale $s=\log(\Lambda/k)$. The plot shows two scaling plateaus described the the Ising (the upper curve) and XY universality classes. For the illustration we chose the quartic coupling $u_0=5$ close to its fixed point value.   } 
\end{center}
\end{figure} 

Leaving aside the Ising-XY crossover we now demonstrate the effect of the anisotropy on the RG flow in the low-$T$ phase.\cite{Leonard_2015} Fig.~5 illustrates the crossover between the scaling behavior controlled by the critical XY (Wilson-Fisher) and the low-$T$ (Nambu-Goldstone) fixed points (both located at $\tilde{\lambda}=0$). We plot the flow of $\tilde{u}$ and $\tilde{Y}$, where the fixed-point behavior is clearly visible. 
\begin{figure}
  \begin{subfigure}[b]{0.23\textwidth}
    \includegraphics[width=\textwidth]{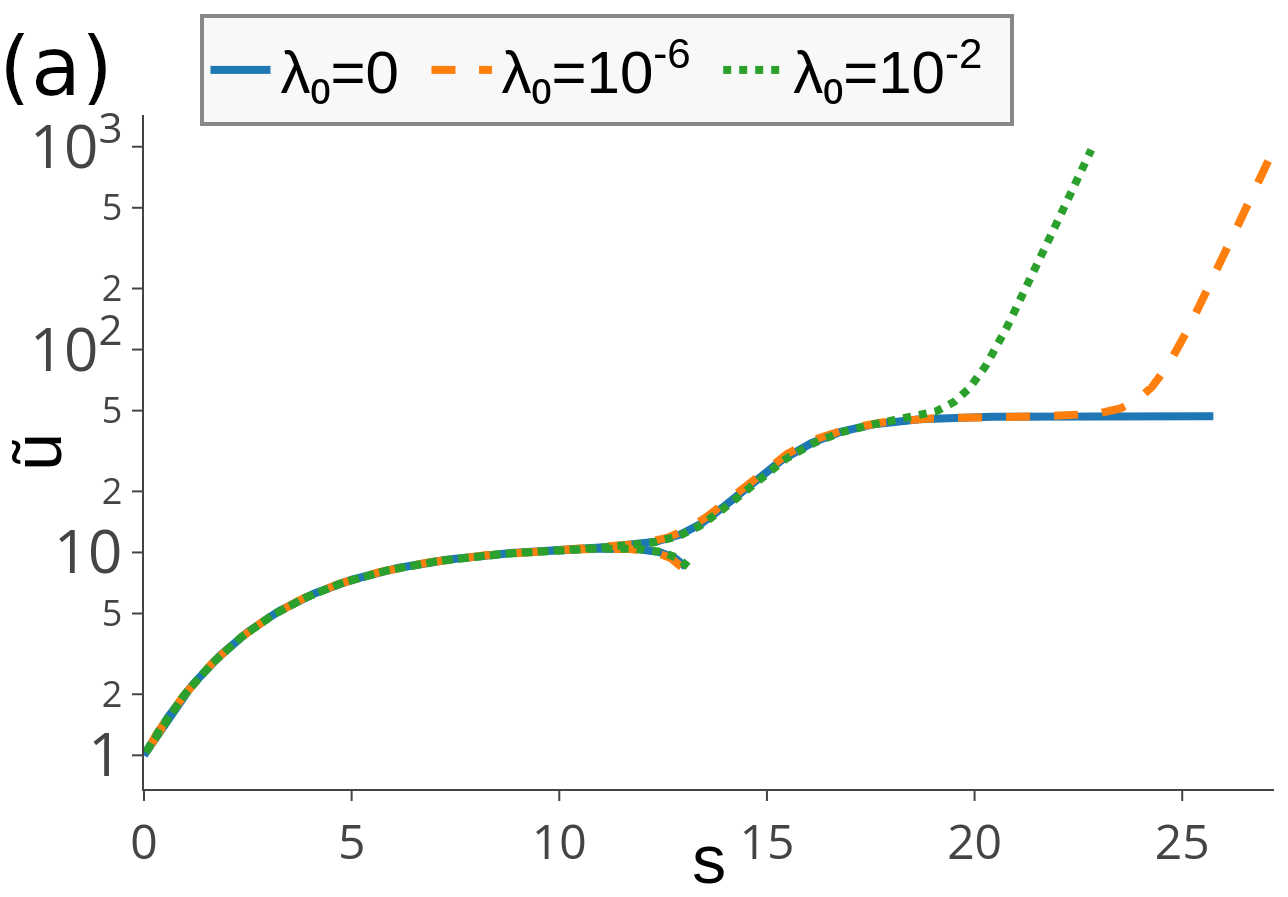}
  %  \caption{Picture 1}
   % \label{fig:1}
  \end{subfigure}
  \begin{subfigure}[b]{0.23\textwidth}
    \includegraphics[width=\textwidth]{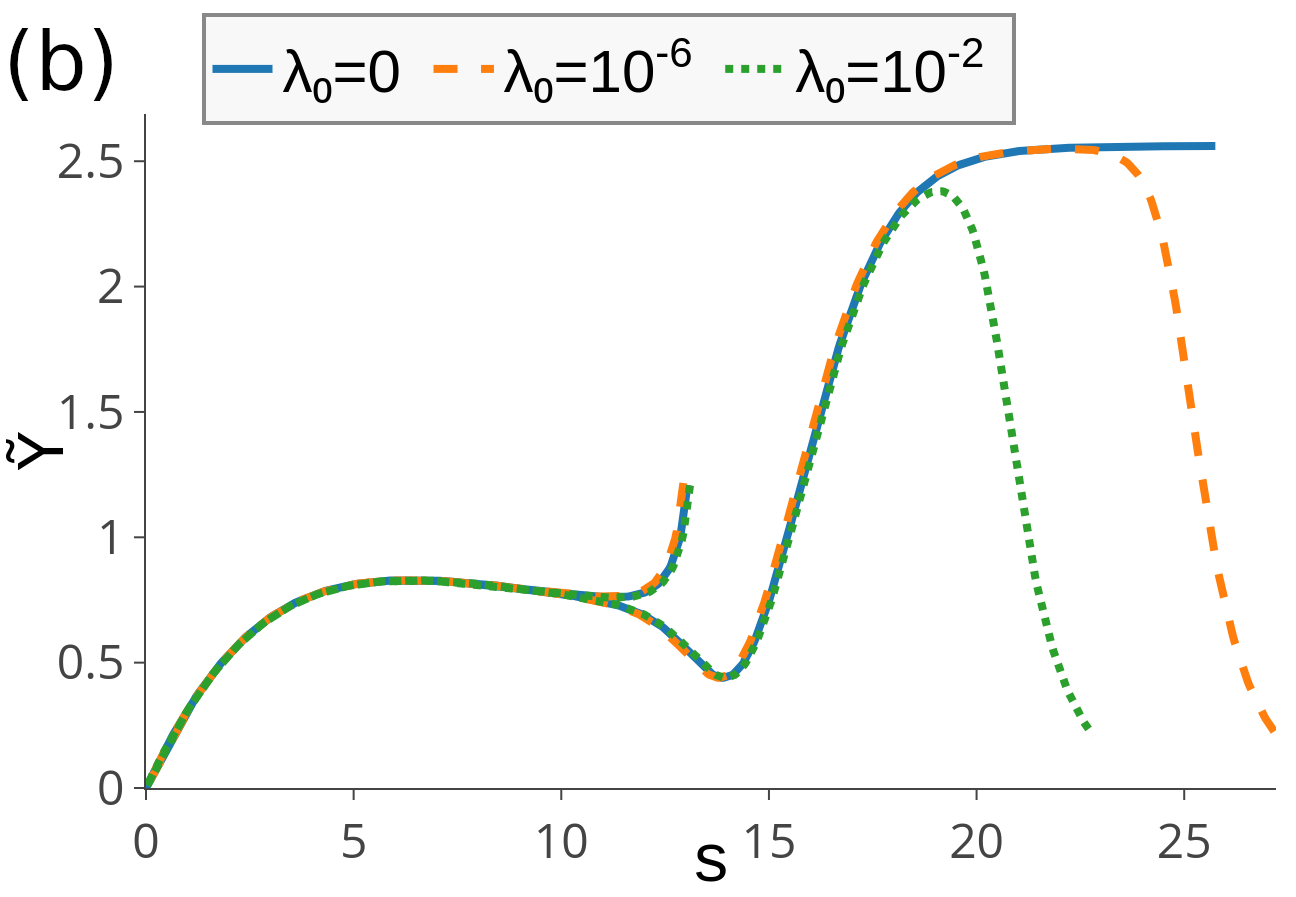}
   % \caption{Picture 2}
    %\label{fig:2}
  \end{subfigure} 
  \caption{(Color online) Illustration of the crossover between the behavior controlled by the critical XY and the low-$T$ fixed points. For an initial condition somewhat below the critical temperature the flow first converges to the vicinity of the Wilson-Fisher fixed point, from which it departs at a scale $k$ of order of the inverse correlation length $\xi^{-1}$ of the longitudinal mode. The subsequent part of the flow is controlled by the $T=0$ fixed point, which, for $\lambda=0$, is never left down to $k=0$. However, if $\lambda\neq 0$, the flow departs from this fixed point at another critical scale $k=\xi'^{-1}$ related to the transverse mode. The presence of these two scales divergent at the phase transition forms the basis of the generic mechanism leading to distinct critical exponents in the low- and high-temperature phases.\cite{Leonard_2015}    }
\end{figure}

For an initial condition somewhat below the critical temperature the flow first converges to the vicinity of the Wilson-Fisher fixed point, from which it departs at a scale $k$ of order of the inverse correlation length $\xi^{-1}$ of the longitudinal mode. The subsequent part of the flow is controlled by the Nambu-Goldstone fixed point, which, for $\lambda=0$, is never left down to $k=0$. However, if $\lambda\neq 0$, the flow departs from this fixed point at another critical scale $k=\xi'^{-1}$ related to the transverse mode. The presence of these two scales divergent at the phase transition is the source of the generic mechanism leading to distinct critical exponents in the low- and high-temperature phases.\cite{Leonard_2015}      

In the next section, we discuss evolution of this picture when dimensionality is changed to $d=2$, where, according to earlier studies, one expects a deformation of the flow diagram involving  appearance of distinct fixed-point lines describing the KT phase and the phase transition at $\lambda\neq 0$. 

\subsection{Results in $d=2$}
We now discuss the results obtained in the simple truncation for $d=2$. We recall\cite{Graeter_1995, Gersdorff_2001, Jakubczyk_2014, Defenu_2017, Jakubczyk_2017_2} that the present approach does not \emph{stricto sensu} capture the KT phase transition, which (due to approximation) is rounded to a very sharp crossover. The correlation length becomes huge (but is still finite) and the KT line of fixed points is visible as a ''quasi-fixed'' point line, where the flow becomes very slow, but ultimately, at very large RG times, ends up in the high-temperature phase. A similar phenomenon occurs, in the present truncation for the fixed-point line at $\lambda>0$. We begin with plotting the phase diagram - see Fig.~6, which may be compared to the case of $d=3$. 
\begin{figure}[ht]
\begin{center}
\label{}
\includegraphics[width=8.5cm]{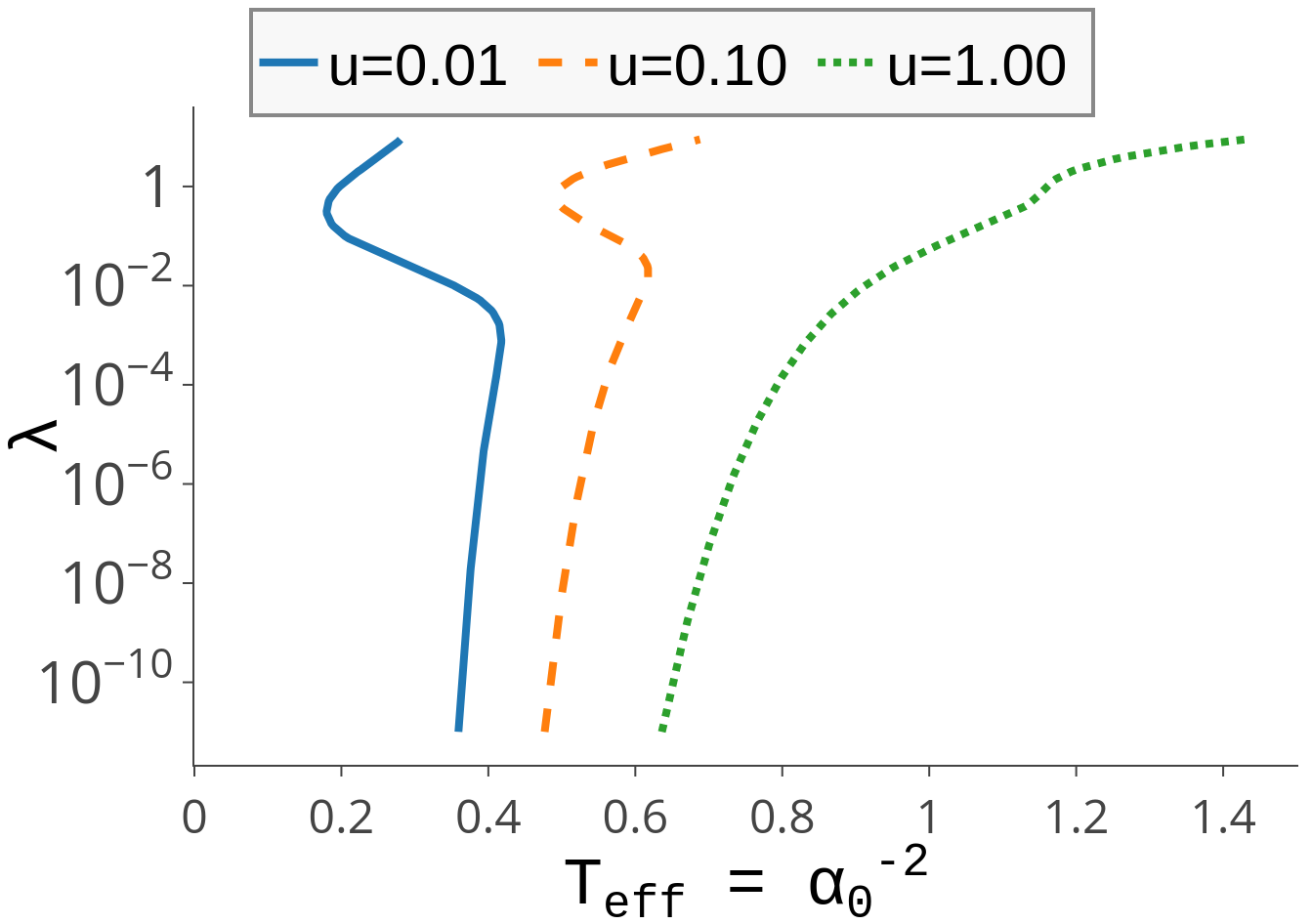}
\caption{(Color online) The critical line $\lambda_{0,u}(\alpha_0)$ computed for a sequence of values of $u_0$ in $d=2$. The sudden drop of $T_{eff}$ occurs precisely at $\lambda_0\approx u_0$, where the transition is in the Ising universality class. The effect is similar to the one observed for $d=3$ (see Fig.~2).  }
\end{center}
\end{figure} 
The behavior of the critical line at $\lambda_0\approx u_0$ is very similar to the case $d=3$ and may be understood as a signature of the Ising fixed point. For small $\lambda$ our result apparently indicates an approach of $T_{eff}$ towards a constant value, which is consistent with the results of Ref.~\onlinecite{Jose_1977} obtained from the Villain model. However, within an alternative (presumably less reliable) approach based on the Migdal transformation, Ref.~\onlinecite{Jose_1977} provided another estimate of the critical line:
 \begin{equation}
    \label{migdal}
    \lambda^c_p(T) \propto e^{-A T^2 e^{B/T}},
\end{equation}  
which is not possible to exclude using our numerical data.  As concerns the flow, a schematic illustration of the picture emergent in our approximation is presented in Fig.~7.
\begin{figure}
  \begin{subfigure}[b]{0.23\textwidth}
    \includegraphics[width=\textwidth]{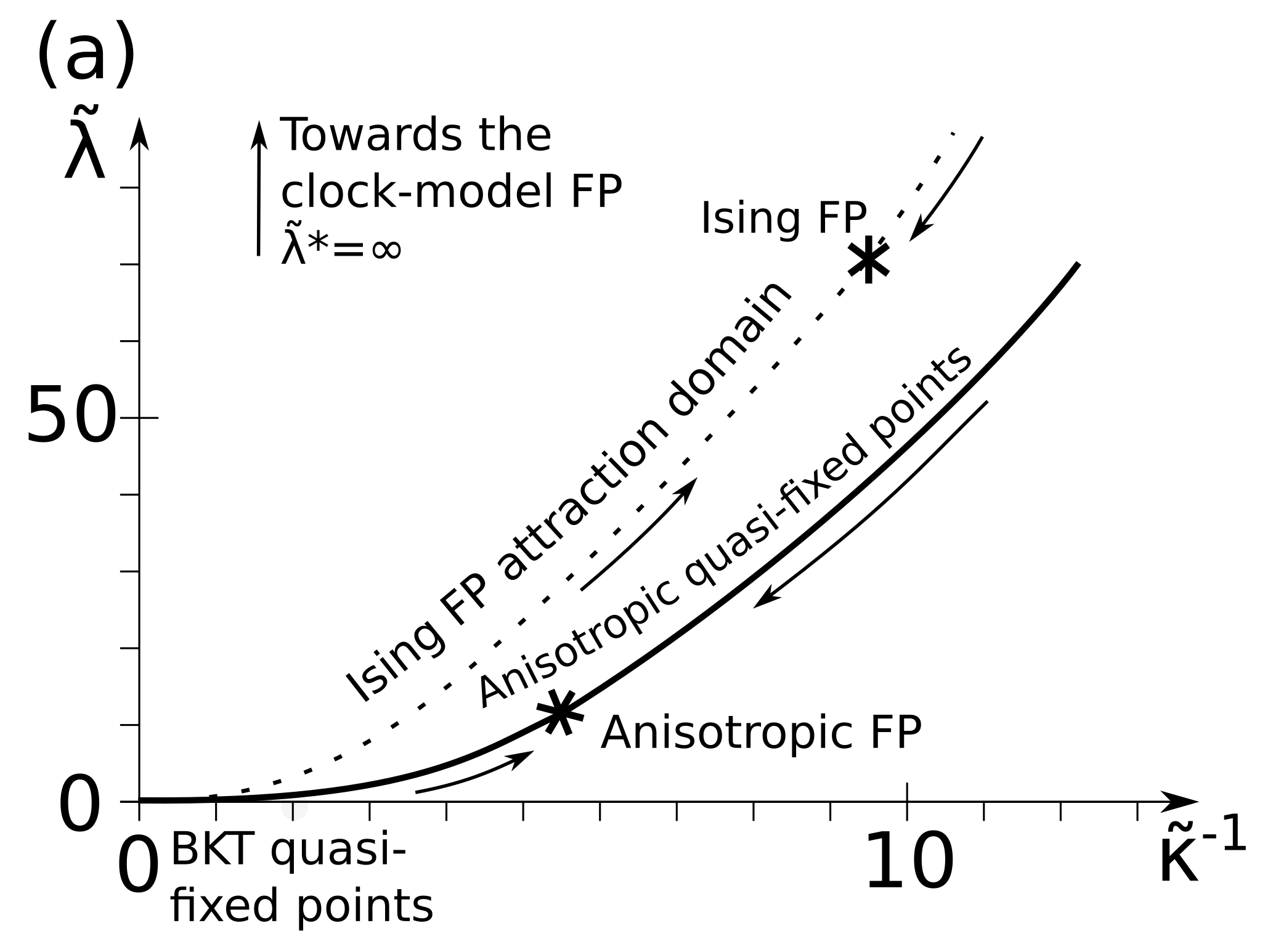}
  %  \caption{Picture 1}
   % \label{fig:1}
  \end{subfigure}
  \begin{subfigure}[b]{0.23\textwidth}
    \includegraphics[width=\textwidth]{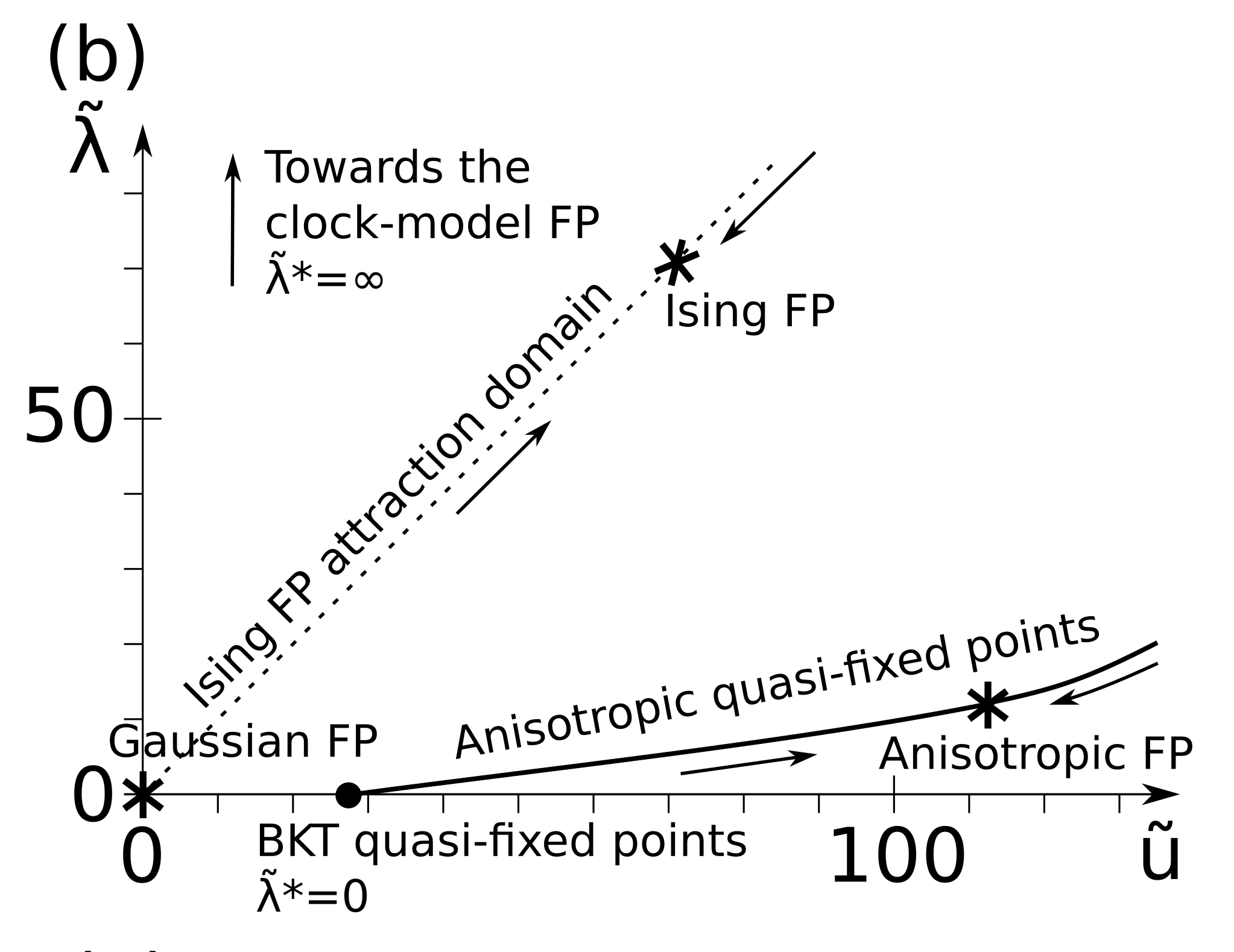}
   % \caption{Picture 2}
    %\label{fig:2}
  \end{subfigure} 
  \caption{A schematic illustration of the picture obtained within the present approximation in $d=2$. The flow diagram features two (quasi)fixed point lines, which, in an exact calculation should turn into exact fixed points. The line located at $\tilde{\lambda}=0$ corresponds to the algebraic KT phase. The line emerging towards large $\lambda$ characterizes the critical points with non-universal exponents. A system tuned to criticality approaches the line and exhibits a very slow flow along it. A single point on this line corresponds to an exact fixed point. The locus $\tilde{\lambda}=\tilde{u}$ is controlled by the Ising fixed point, completely detached from the (quasi)fixed point line. }
\end{figure}
In addition to the KT (quasi)fixed point line at $\tilde{\lambda}=0$ we identify a similar line extending towards large $\lambda$. A system tuned to the critical point (at some $\lambda_0>0$)  approaches the line, and exhibits a regime of very slow flow along the line. We note that the entire (quasi)fixed line is located in the regime $\tilde{\lambda}<\tilde{u}$. The regime $\lambda_0>u_0$ is beyond the scope of the present study. The behavior of the system for the special choice of the initial condition $\lambda_0=u_0$ (and $\alpha_0$ tuned to the critical value) corresponds to the Ising universality class in full analogy to the case of $d=3$.  However, for small $\lambda_0$ an intermediate $KT$ scaling sets in so that the flow first proceeds along the KT line and crosses over to the scaling controlled by the line at $\tilde\lambda$ only at low scales. This crossover scale diverges for small $\lambda_0$. This behavior is well illustrated in Fig.~8, where we plot $\tilde{u}$ versus the cutoff scale. 
\begin{figure}[ht]
\begin{center}
\label{}
\includegraphics[width=8.5cm]{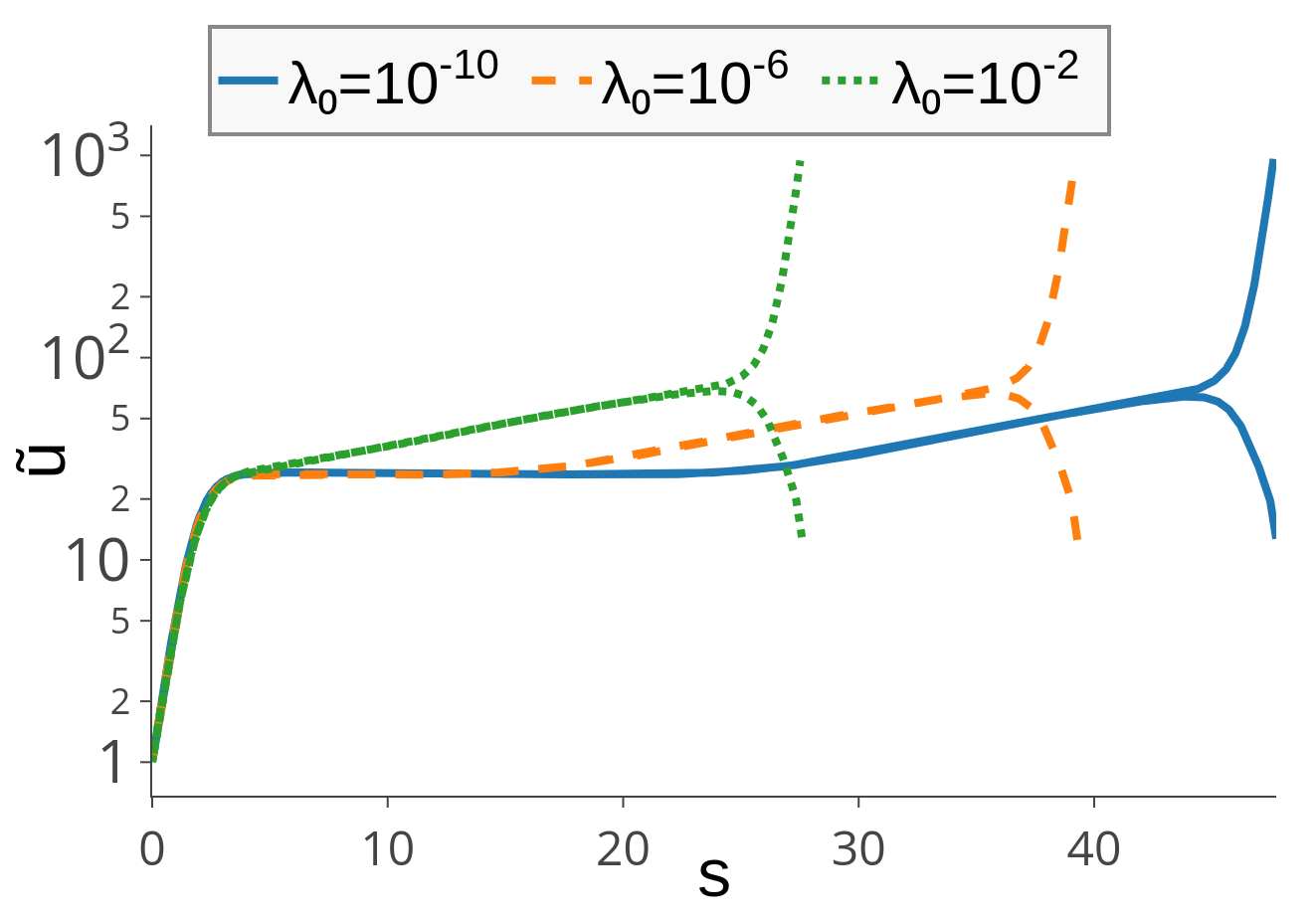}
\caption{(Color online)  Flow of $\tilde{u}$ for a sequence of values of the initial anisotropy coupling. For sufficiently low $\lambda_0$ (e.g. the lowest, blue curve) the flow rapidly approaches the KT (quasi)fixed point line and remains in its vicinity for a substantial RG time. Subsequently it crosses over to the behavior governed by the finite $\tilde\lambda$ (quasi)fixed point line from which it ultimately departs due to numerical limitations. The crossover scale diverges for $\lambda_0\to 0$. For sufficiently large $\lambda_0$ the KT scaling is not visible and the flow immediately runs towards the finite $\tilde\lambda$ fixed-point line (see e.g. the highest, green curve).  }
\end{center}
\end{figure} 
In an exact calculation we expect exact fixed-point behavior along the line at $\tilde\lambda>0$ in accord with Ref~\onlinecite{Jose_1977}. Note however, that  the existence of this line is fully established only for small anisotropies. The other worthwhile observation is that the (quasi)fixed point line obtained by us is located for $\tilde\lambda$ significantly smaller than $\tilde u$.    
The two (quasi)fixed point lines are well identified in the flow diagram in the $(\tilde\lambda, \tilde{\kappa}^{-1}, \tilde u)$ space, and we present the projections in Figs.~9 and 10. 
\begin{figure}
  \begin{subfigure}[b]{0.23\textwidth}
    \includegraphics[width=\textwidth]{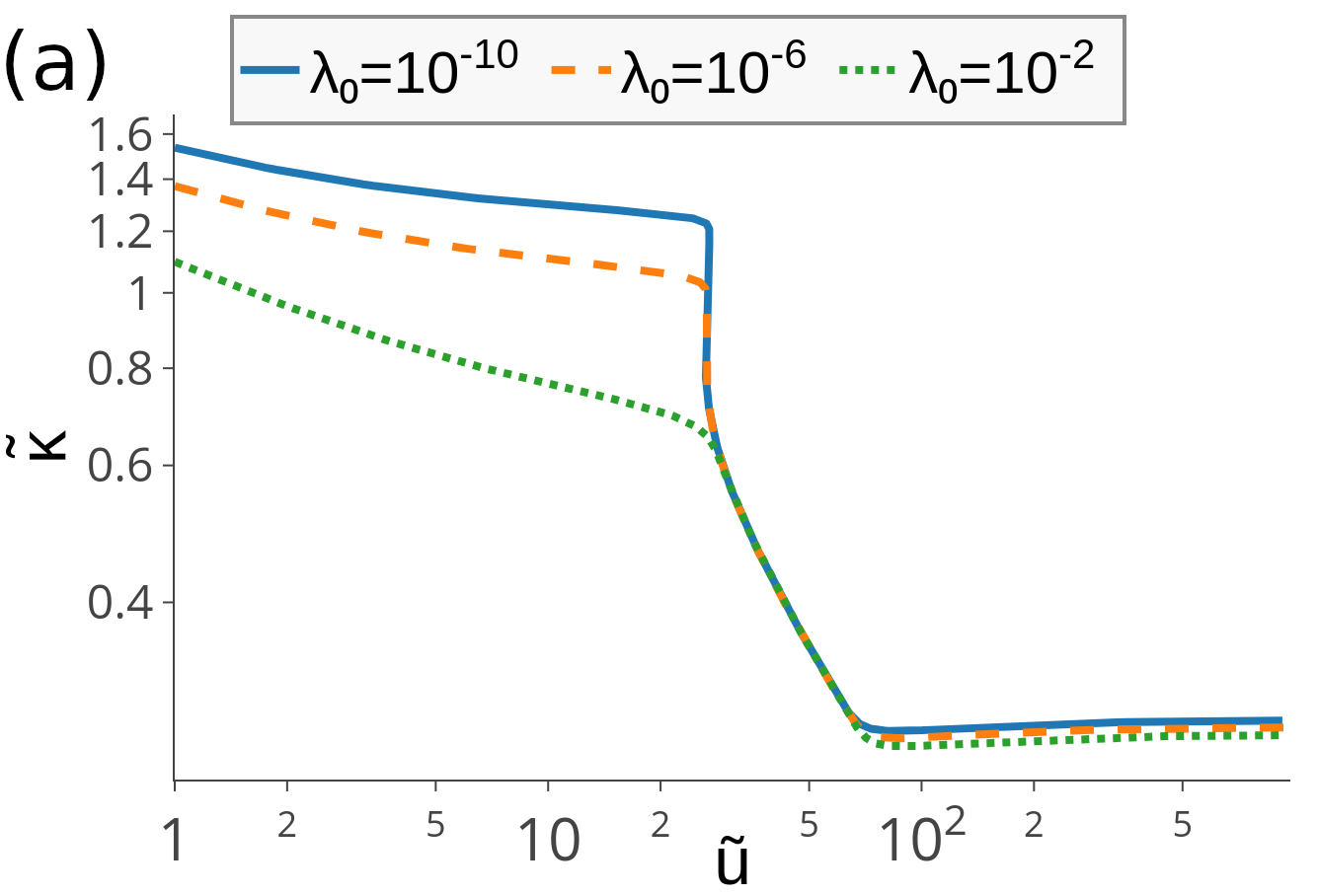}
  %  \caption{Picture 1}
   % \label{fig:1}
  \end{subfigure}
  \begin{subfigure}[b]{0.23\textwidth}
    \includegraphics[width=\textwidth]{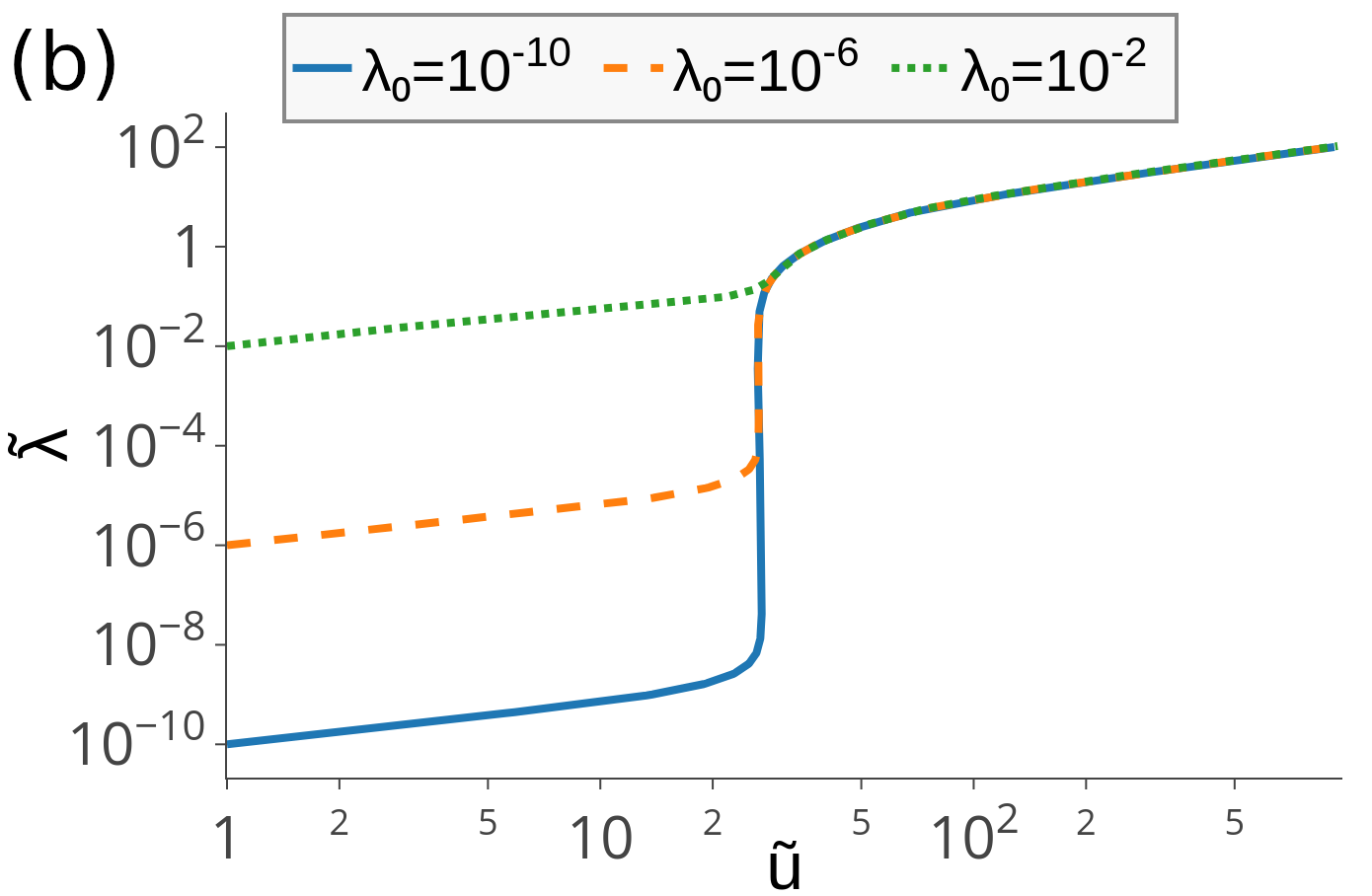}
   % \caption{Picture 2}
    %\label{fig:2}
  \end{subfigure} 
  \caption{(Color online) Projections of the numerically obtained RG flow diagrams demonstrating the (quasi)fixed point lines. (Compare to Fig.~8.) The (almost) vertical locus represents the KT fixed point line, which crosses over to the tilted line followed by all the plotted curves, corresponding to the finite $\tilde\lambda$ fixed-point line. In the final part of the flow the curves depart from the (quasi)fixed point line and go into the low-$T$ phase.   }
\end{figure}
Generically, all the critical flows converge to one universal line in the $(\tilde\lambda, \tilde{\kappa}^{-1}, \tilde u)$ space, which they follow up to a point determined by numerical accuracy of tuning to the critical point and integrating the flow. The crossover scale between the KT and critical (finite $\tilde{\lambda}$) behavior diverges for $\lambda_0\to 0$. It is not hard to imagine pushing it to values way below the scale controlled by the system size in simulations. This explains the presence of the (apparent) KT phase in Monte-Carlo data.
\begin{figure}[ht]
\begin{center}
\label{}
\includegraphics[width=8.5cm]{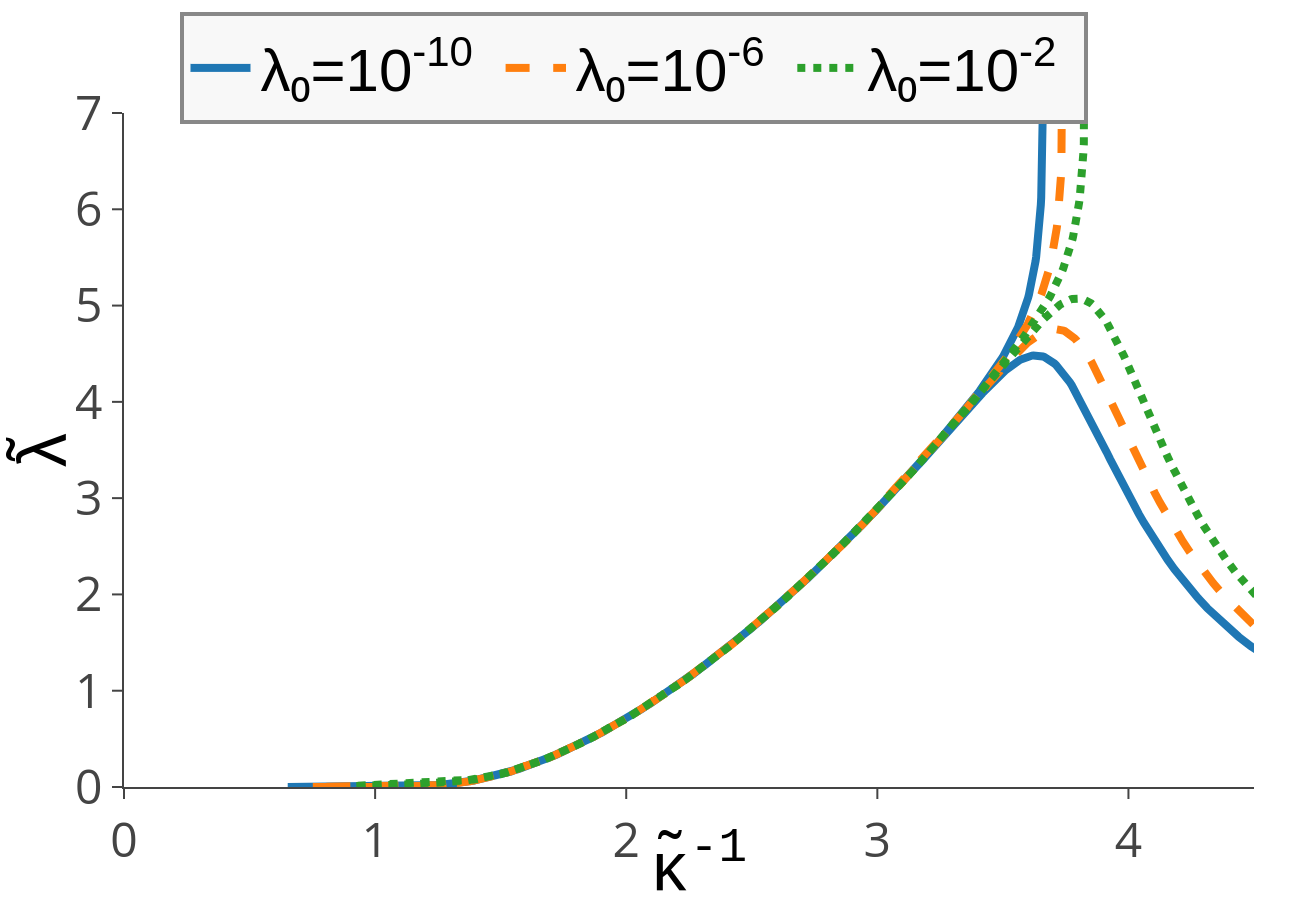}
\caption{(Color online) Projection of the RG flows from Fig.~9 on the $(\tilde{\kappa}^{-1}, \tilde{\lambda})$ plane.  }
\end{center}
\end{figure} 

We finally discuss the critical exponents. The values obtained at the present approximation level may serve only as crude estimate. The applied approximation level allows however for observing signatures of nonuniversality. This is demonstrated by plotting the transverse and longitudinal susceptibility exponents as function of anisotropy in Fig.~11. The obtained values are (almost) equal and vary with $\lambda_0$. This is contrasted to the results obtained in $d=3$, where the exponents are (as expected) different from each other and independent of $\lambda_0$.  
\begin{figure}
  \begin{subfigure}[b]{0.23\textwidth}
    \includegraphics[width=\textwidth]{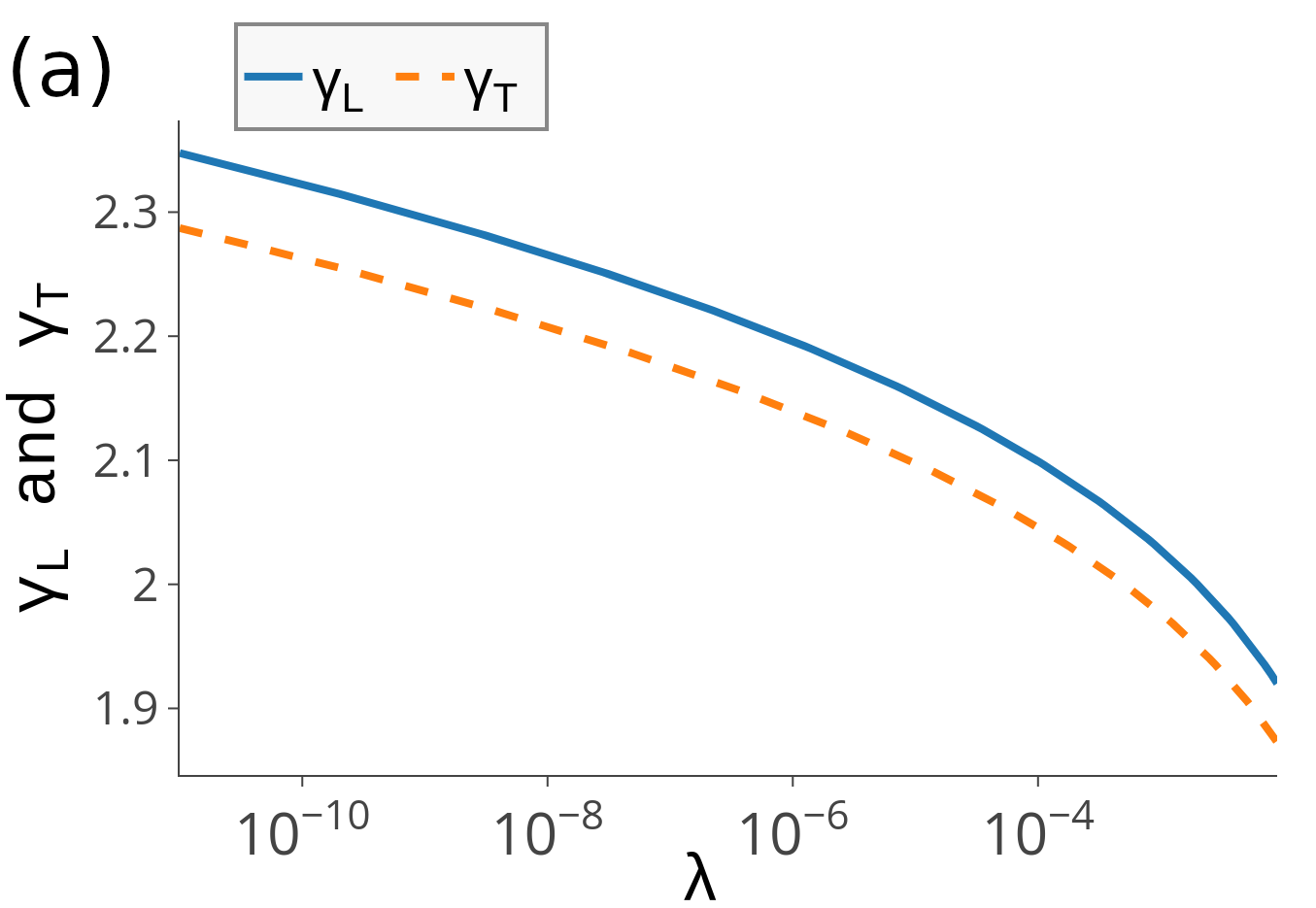}
  %  \caption{Picture 1}
   % \label{fig:1}
  \end{subfigure}
  \begin{subfigure}[b]{0.23\textwidth}
    \includegraphics[width=\textwidth]{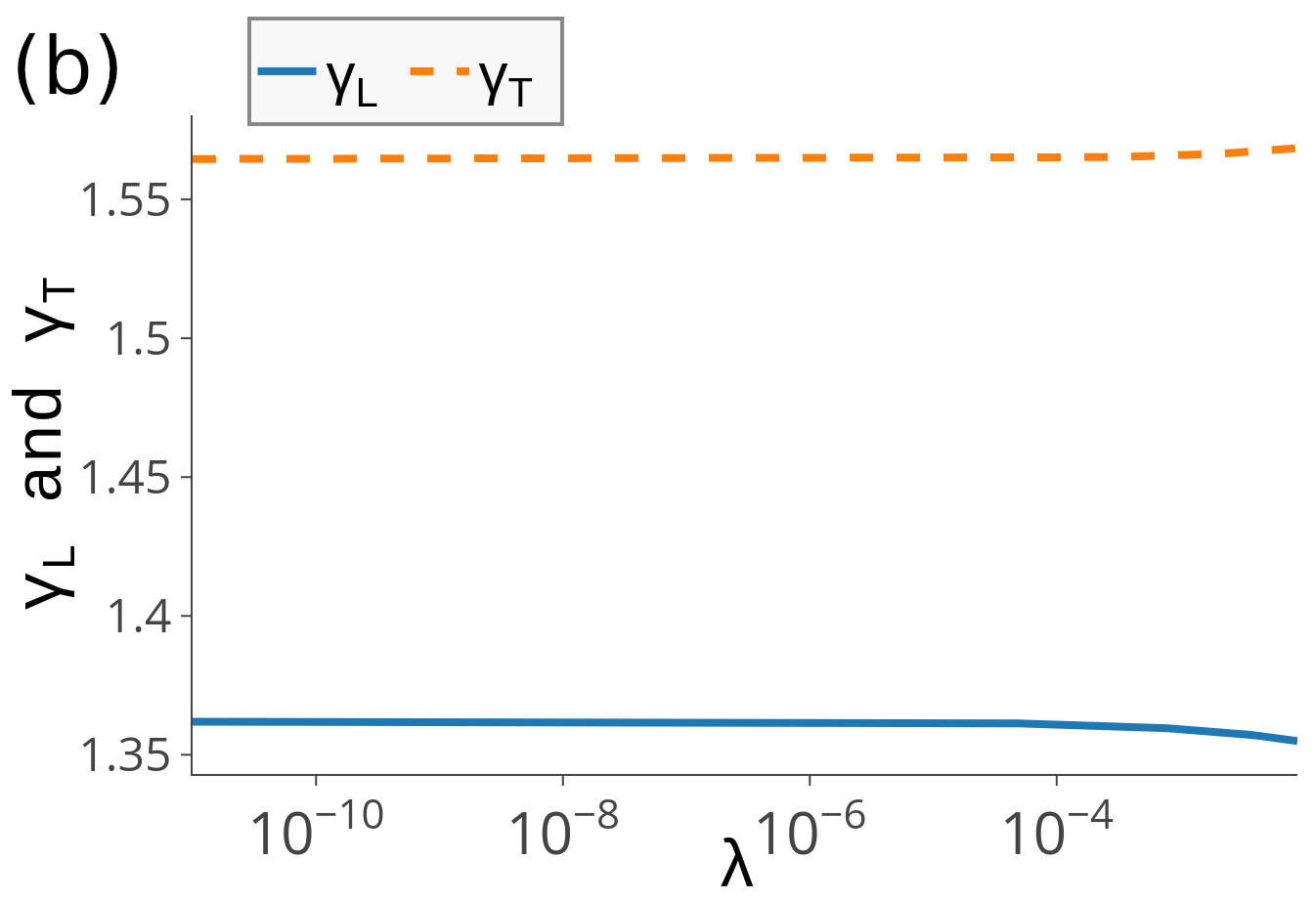}
   % \caption{Picture 2}
    %\label{fig:2}
  \end{subfigure} 
  \caption{Transverse and longitudinal susceptibility exponents as function of anisotropy $\lambda_0$ for $d=2$ (left panel) and $d=3$ (right panel). In $d=2$ the obtained exponents are (almost) equal, but show dependence on $\lambda_0$. In $d=3$ the two exponents are different, but universal. The range of $\lambda_0$ is chosen small to avoid any possible interference with the Ising fixed point.   }
\end{figure} 

One feature concerning the limit $\lambda_0\to 0$ in $d=2$ cannot be accounted for by the present functional RG truncation. This is related to the essential singularity of the correlation length of the pure $O(2)$ model, which is not captured by the applied approximation and requires going at least to 2nd order in complete derivative expansion.\cite{Jakubczyk_2014} This deficiency has impact on the system also for $\lambda_0\neq 0$. Indeed, reasoning by continuity the nonuniversal correlation length exponent $\nu$ should diverge for vanishing $\lambda_0$. By scaling laws this implies also the divergence of the susceptibility and order parameter $\beta$ exponents. This divergence, predicted by Ref.~\onlinecite{Jose_1977} (see also Ref.~\onlinecite{Kadanoff_1977}) is not captured by the present simple truncation, and, in the present framework, requires a higher-order treatment. We leave this, together with an accurate resolution of the exponents (including the scaling exponent of the anisotropy coupling $\lambda$) to future work. 
%\subsection{Results in $d\in (2,3)$}
%\section{Beyond the simple truncation}
\section{Summary and perspective}
We employed a simple truncation of the nonperturbative renormalization group to analyze the impact of $\mathbb{Z}_4$-symmetric perturbations on the critical behavior of the $O(2)$ model in dimensionality $d=3$ and $d=2$. This allowed us to treat the two rather different situations in a unified framework and resolve the relatively complex crossover behavior arising due to the interplay of distinct RG fixed points. Both for $d=2$ and $d=3$ in the parameter space there exists a (one-dimensional) domain of attraction of the Ising fixed point, whose presence manifests itself in specific values of the critical exponents, but also via an abrupt variation of the critical temperature as function of the anisotropy coupling. In three dimensions, apart from the special situation $\lambda_0=u_0$, the flow at asymptotically low scales is controlled by the Wilson-Fisher fixed point. However, due to the presence of the dangerously irrelevant anisotropy coupling, the Goldstone mode acquires a gap and the critical exponents differ depending on the side from which the phase transition is approached.\cite{Leonard_2015} This behavior may also be understood by realizing the existence of two distinct divergent scales controlling the flow in regimes corresponding to the vicinity of the critical (XY) and the low-$T$ fixed points. Therefore, the complete picture in $d=3$ involves the interplay of three fixed points and the rich related crossover behavior. 

In $d=2$, in addition to the Kosterlitz-Thouless fixed point line and the abovementioned Ising fixed point, there exists a separate line of fixed points emerging towards the regime of large anisotropy couplings. Within the present treatment (in an analogy with the Kosterlitz-Thouless phase) this is captured approximately by a flow regime characterized by very slow running of the couplings in the form ot the quasi-fixed points. The RG flow finds itself between the isolated Ising fixed point and the two lines. For small anisotropy coupling the critical trajectories follow the close vicinity of the KT line down to an RG scale where they cross over to the critical fixed-point line. This crossover scale diverges for vanishing anisotropies so that the flow becomes dominated by the KT line down to extremely low RG scales. This picture gives a natural explanation of the KT-like behavior observed in Monte-Carlo simulations of the lattice $XY$ model at sufficiently small anisotropies.  

There are natural extensions of this work worth investigating in near future. These involve a more accurate, functional parametrization of the flowing effective action, avoiding expansion of the flowing effective potential $U_k(\rho, \tau)$ in the invariants $\rho$ and $\tau$. In $d=2$ this would allow for capturing the KT transition (including the essential singularity of the correlation length) and therefore an accurate resolution of the limit $\lambda\to 0$. We expect that such a truncation would in fact give a very accurate picture at $\lambda>0$. This is because vortices, which are hard to capture by the present approach, are actually not relevant for a description of physics at finite anisotropies. The other observation is that the existence of the line of fixed points at $\lambda\neq 0$ is firmly established only in the limit $\lambda\to 0$. The present framework (at functional level) is by no means restricted to this regime. In particular, it might be very interesting to investigate the case $\lambda_0>u_0$ and the emerging relation to the $p$-state clock models.\cite{Lapilli_2006}

\begin{acknowledgments}
We thank Bertrand Delamotte, Maxym Dudka, Nicolas Dupuis, and Marek Napi\'{o}rkowski for useful discussions. We acknowledge support from the Polish National Science Center via 
2014/15/B/ST3/02212 and 2017/26/E/ST3/00211. 
\end{acknowledgments}

\section*{Appendix} 
Here we quote the flow equations for the $Z$-factors $Z_1$ and $Z_2$ corresponding to the longitudinal and transverse modes respectively. We introduce $R^{(n,m)}(y) := \frac{d^{n+m} R_k(y)}{dy^n dk^m}$. The flow equations read:
\begin{widetext} 
\begin{align*}
\partial_k{Z_{1}} &= \frac{ Y}{ u} \int_{\bm{q}}G_1^2(\bm{q})2U(\bm{q})R^{(0,1)} 
+ \int_{\bm{q}} G_1^3(\bm{q})\alpha^2\Bigg(\frac{1}{2} (u+2U(\bm{q}))^2(R^{(1,1)} + \frac{2}{d}\bm{q}^2R^{(2,1)}) + 2 R^{(0,1)} Y(2u+4U(\bm{q})+ \frac{2}{d} \bm q^2 Y)  \\
&\hspace{5.7cm} + 2 \frac{2}{d} \bm{q}^2 Y (u+2U(\bm{q})) R^{(1,1)} \Bigg)  \\
&- \int_{\bm{q}} G_1^4(\bm{q}) \alpha^2 \Bigg(2 \frac{2}{d} \bm{q}^2 (u+2U(\bm{q}))^2 (Z_1 + R^{(1,0)}) R^{(1,1)}  
 + \frac{3}{2} (u+2U(\bm{q}))^2 (Z_1+R^{(1,0)}+\frac{2}{d}\bm{q}^2R^{(2,0)}) R^{(0,1)}  \\
&\hspace{2.3cm} + 6\frac{2}{d} \bm{q}^2 (u+2U(\bm{q})) (Z_1+R^{(1,0)}) R^{(0,1)}\Bigg)  \\
&+ \int_{\bm{q}} G_1^5(\bm{q}) 4 \alpha^2 \frac{2}{d}\bm{q}^2 (u+2U(\bm{q}))^2 \left(Z_1 + R^{(1,0)}\right)^2 R^{(0,1)}  
+ \frac{ Y}{ u} \int_{\bm{q}} G_2^2(\bm{q}) (u+2\lambda)R^{(0,1)}  \\
&+ \int_{\bm{q}} G_2^3(\bm{q})\alpha^2\Bigg(\frac{1}{2} (u+2\lambda)^2 \left(R^{(1,1)} + \frac{2}{d}\bm{q}^2R^{(2,1)}\right) + 2  Y(u+2\lambda)R^{(0,1)} \Bigg)  \\
&- \int_{\bm{q}} G_2^4(\bm{q}) \alpha^2 (u+2\lambda)^2 \Bigg(2\frac{2}{d} \bm{q}^2 (Z_2 + R^{(1,0)}) R^{(1,1)}  
+ \frac{3}{2} (Z_2+R^{(1,0)}+\frac{2}{d}\bm{q}^2R^{(2,0)}) R^{(0,1)}\Bigg)  \\
&+ \int_{\bm{q}} G_2^5(\bm{q}) 4 \alpha^2 \frac{2}{d}\bm{q}^2 (u+2\lambda)^2 \left(Z_2 + R^{(1,0)}\right)^2 R^{(0,1)}
\end{align*}
and
\begin{align*}
\partial_k{Z_{2}} &= - \int_{\bm{q}} Y R^{(0,1)}(\bm{q}^2)G_{2}^2(\bm{q}) 
+ \int_{\bm{q}} 6 \alpha^2 \frac{2}{d}\bm{q}^2 (2\lambda + U(\bm{q}))^2 R^{(0,1)}(\bm{q}^2) \left(Z+R^{(1,0)}(\bm{q}^2)\right)^2 G_2^4(\bm{q}) G_1(\bm{q})  \\
&+ \int_{\bm{q}} 2 \alpha^2 \frac{2}{d}\bm{q}^2 (2\lambda + U(\bm{q}))^2 R^{(0,1)}(\bm{q}^2) \left(Z+R^{(1,0)}(\bm{q}^2)\right)^2 G_2^3(\bm{q}) G_1^2(\bm{q})  \\
&- \int_{\bm{q}} 4 \alpha^2 \frac{2}{d}\bm{q}^2 (2\lambda + U(\bm{q}))^2 R^{(1,1)}(\bm{q}^2) \left(Z+R^{(1,0)}(\bm{q}^2)\right) G_2^3(\bm{q}) G_1(\bm{q})  \\
&- \int_{\bm{q}} 2 \alpha^2 (2\lambda + U(\bm{q}))^2 R^{(0,1)}(\bm{q}^2) \left(Z+R^{(1,0)}(\bm{q}^2)+\frac{2}{d}\bm{q}^2 R^{(2,0)}(\bm{q}^2)\right) G_2^3(\bm{q}) G_1(\bm{q})  \\
&- \int_{\bm{q}} \alpha^2 (2\lambda + U(\bm{q}))^2 R^{(0,1)}(\bm{q}^2) \left(Z+R^{(1,0)}(\bm{q}^2)+\frac{2}{d}\bm{q}^2 R^{(2,0)}(\bm{q}^2)\right) G_2^2(\bm{q}) G_1^2(\bm{q})  \\
&+ \int_{\bm{q}} \alpha^2 (2\lambda + U(\bm{q}))^2 (R^{(1,1)}(\bm{q}^2) + \frac{2}{d}\bm{q}^2 R^{(2,1)}(\bm{q}^2)) G_2^2(\bm{q}) G_1(\bm{q})\;.
\end{align*}
\end{widetext}

\bibliography{/Users/pjak/Desktop/refs}{} 
%\bibliography{/Users/Pawel/Desktop/Paper_with_Piotr/Version_PJ/refs}{} 
\bibliographystyle{apsrev4-1}

\end{document}